\documentclass[preprintnumbers,superscriptaddress,pra,showkeys]{revtex4}
\usepackage{amsfonts}
\usepackage{amssymb}
\usepackage{amsmath}
\usepackage{epsfig}
\usepackage{graphicx}
\usepackage{color}

\input{tcilatex}
\begin{document}

\title{A new discrete calculus of variations and its applications in
statistical physics}
\author{Q. H. Liu}
\email{quanhuiliu@gmail.com}
\affiliation{School for Theoretical Physics, School of Physics and Electronics, Hunan
University, Changsha 410082, China}
\date{\today }

\begin{abstract}
For a discrete function $f\left( x\right) $ on a discrete set, the finite
difference can be either forward and backward. However, we observe that if $%
f\left( x\right) $ is a sum of two functions $f\left( x\right) =f_{1}\left(
x\right) +f_{2}\left( x\right) $ defined on the discrete set, the first
order difference of $\Delta f\left( x\right) $ is equivocal for we may have $%
\Delta ^{f}f_{1}\left( x\right) +\Delta ^{b}f_{2}\left( x\right) $ where $%
\Delta ^{f}$ and $\Delta ^{b}$ denotes the forward and backward difference
respectively. Thus, the first order variation equation for this function $%
f\left( x\right) $ gives many solutions which include both true and false
one. A proper formalism of the discrete calculus of variations is proposed
to single out the true one by examination of the second order variations,
and is capable of yielding the exact form of the distributions for
Boltzmann, Bose and Fermi system without requiring the numbers of particle
to be infinitely large. The advantage and peculiarity of our formalism are
explicitly illustrated by the derivation of the Bose distribution.
\end{abstract}

\author{}
\maketitle

\emph{Introduction} The method of\emph{\ most probable distribution }(MPD)%
\emph{\ }is the most common way used to derive various statistical
distributions in mathematics, physics, chemistry, materials science and
computational science, etc. The original data are usually discrete rather
than continuous, so we must be able to deal with the differences, difference
quotients and sums of discrete functions, instead of the differentials,
derivatives and integrations of the continuous functions. However, we
immediately run into a problem: For a discrete function that has extremals,
the difference can be forward, backward, central, and even more complicated
combinations of these differences, so that the first order variation
equation leads to many solutions which can not be all true. By the true
solution, we mean that it is realistic or physical. There has been no
satisfactory procedure to single out the true one in its own right. Thus the
derivation of the exact form of the distribution functions for Boltzmann,
Bose and Fermi system obtained from the MPD has been an open problem for
nearly one century. \cite{chem,PLA} In present work, I\ report a discrete
calculus of variations to resolve this problem.

\emph{A new discrete calculus of variations }Let us first define \emph{an} 
\emph{individual function} (IF) $f\left( x\right) $ of variable $x$ on an
interval $C$, which amounts to that the exponential of the function $f\left(
x\right) $ as $\exp f\left( x\right) $ can not be further factorized. We can
therefore define the sum of two IFs $f\left( x\right) =f_{1}\left( x\right)
+f_{2}\left( x\right) $, and the sum of many IFs. Secondly we define the 
\emph{unidirectional differences }of the IF\emph{\ }$f\left( x\right) $,
which means that when taking the finite differences of different orders for
the function\emph{\ }$f\left( x\right) $, we take \emph{forward or backward}
differences in all first, second, and higher order variation. For instance,
the \emph{forward directional differences} of the IF\emph{\ }$f\left(
x\right) $ are $\left( \Delta ^{f}\right) f\left( x\right) =f\left(
x+h\right) -f\left( x\right) $, ($h\succ 0$), $\left( \Delta ^{f}\right)
^{2}f\left( x\right) =f\left( x+2h\right) -f\left( x+h\right) -\left(
f\left( x+h\right) -f\left( x\right) \right) =f\left( x+2h\right) -2f\left(
x+h\right) +f\left( x\right) $, and so forth; and the \emph{backward
unidirectional differences} of the IF\emph{\ }$f\left( x\right) $ are $%
\left( \Delta ^{b}\right) f\left( x\right) =f\left( x\right) -f\left(
x-k\right) $, ($k\succ 0$), $\left( \Delta ^{b}\right) ^{2}f\left( x\right)
=f\left( x\right) -f\left( x-k\right) -\left( f\left( x-k\right) -f\left(
x-2k\right) \right) =f\left( x\right) -2f\left( x-k\right) +f\left(
x-2k\right) $, and so forth. Thus, the \emph{forward directional difference
quotients }are $\left( \Delta ^{f}\right) f\left( x\right) /h$, $\left(
\Delta ^{f}\right) ^{2}f\left( x\right) /h^{2}$, ...; and similarly, the 
\emph{backward directional difference quotients }are $\left( \Delta
^{b}\right) f\left( x\right) /k$, $\left( \Delta ^{b}\right) ^{2}f\left(
x\right) /h^{2}$, .... Specially for $h=1$, we have $\left( \Delta
^{f}\right) ^{i}f\left( x\right) /h=\left( \Delta ^{f}\right) ^{i}f\left(
x\right) $, ($i=1,2,3,...$); and similarly for $k=1$, we have $\left( \Delta
^{b}\right) ^{i}f\left( x\right) /k=\left( \Delta ^{b}\right) ^{i}f\left(
x\right) $. The \emph{unidirectional differences} are similar to the
treatment of the parameter $\lambda \in 
\mathbb{R}
$ in the usual calculus of variation in which for the given variational
parameter $\lambda $, we must keep using this parameter $\lambda $ in all
orders of variations. \cite{arnold} However, the different point of the new
procedure is that, once $f\left( x\right) =f_{1}\left( x\right) +f_{2}\left(
x\right) $, $\Delta f$ does mean four combinations: $\left( \Delta
^{f}\right) f_{1}\left( x\right) +\left( \Delta ^{f}\right) f_{2}\left(
x\right) $, $\left( \Delta ^{f}\right) f_{1}\left( x\right) +\left( \Delta
^{b}\right) f_{2}\left( x\right) $, $\left( \Delta ^{b}\right) f_{1}\left(
x\right) +\left( \Delta ^{f}\right) f_{2}\left( x\right) $, and $\left(
\Delta ^{b}\right) f_{1}\left( x\right) +\left( \Delta ^{b}\right)
f_{2}\left( x\right) $, which can be simply named by 1f2f, 1f2b, 1b2f and
1b2b, respectively. Moreover, the second order variations are must be taken 
\emph{unidirectionally}, so $\Delta ^{2}f$ for the combination 1f2f, 1f2b,
1b2f and 1b2b gives, $\left( \Delta ^{f}\right) ^{2}f_{1}\left( x\right)
+\left( \Delta ^{f}\right) ^{2}f_{2}\left( x\right) $, $\left( \Delta
^{f}\right) ^{2}f_{1}\left( x\right) +\left( \Delta ^{b}\right)
^{2}f_{2}\left( x\right) $, $\left( \Delta ^{b}\right) ^{2}f_{1}\left(
x\right) +\left( \Delta ^{f}\right) ^{2}f_{2}\left( x\right) $, and $\left(
\Delta ^{b}\right) ^{2}f_{1}\left( x\right) +\left( \Delta ^{b}\right)
^{2}f_{2}\left( x\right) $, respectively.

The problem is formulated in the following. We at first deal with a discrete
IF $\Psi \left( n\right) $ defined, for convenience, on the interval of
semi-positive integers $n\in 
\mathbb{Z}
^{+}$, which has the local maxima and minima, subject to some equality
constraints $\mathbf{\psi =}\left( \psi _{1},\psi _{2},\psi _{3},...\right) =%
\mathbf{0}$. For finding the local maxima and minima of the function $\Psi
\left( n\right) $, we construct a functional $\Phi $,%
\begin{equation}
\Phi =\Psi \left\{ n\right\} +\mathbf{\alpha \cdot \psi ,}  \label{Theorem}
\end{equation}%
where $\mathbf{\alpha =}\left( \alpha _{1},\alpha _{2},\alpha
_{3},...\right) $ are Lagrange multipliers each of which $\alpha _{i}$
companies a constraint condition $\psi _{i}=0$. The local maxima and minima
satisfy, 
\begin{equation}
\delta \Phi =0.  \label{variation}
\end{equation}%
Since the discreteness of the function $\Psi \left( n\right) $, the smallest
finite change of $n$ is $\Delta n=1$, and we have accordingly two
differences (or difference quotients) of the function $\Psi \left\{
n\right\} $ in the following:\ $\left( \delta ^{f}\right) \Psi \left\{
n\right\} /\delta n=\Psi \left\{ n+1\right\} -\Psi \left\{ n\right\} $ $%
=\left( \Delta ^{f}\right) \Psi \left\{ n\right\} $ for the forward
difference and $\left( \delta ^{b}\right) \Psi \left\{ n\right\} /\delta
n=\Psi \left\{ n\right\} -\Psi \left\{ n-1\right\} =\left( \Delta
^{b}\right) \Psi \left\{ n\right\} $ for the backward difference. In
consequence, each of two relations $n^{\mu }=n^{\mu }\left( \mathbf{\alpha }%
\right) $ ($\mu =1,2$) solves the equation (\ref{variation}), constituting
the solution pool which contains both spurious and true one. Then how to
single out the true one?

Assuming that all relations $n^{\mu }=n^{\mu }\left( \mathbf{\alpha }\right) 
$ in the solution pool have different values of the second order differences
in the sense of the \emph{unidirectional differences }of $\Psi \left\{
n\right\} $, and further assuming that $n^{1}=n^{1}\left( \mathbf{\alpha }%
\right) $ is obtained from the forward difference variation as $\left(
\Delta ^{f}\right) \Psi \left\{ n^{1}\right\} +\mathbf{\alpha \cdot }\Delta
^{f}\mathbf{\psi }\left\{ n^{1}\right\} =0$, the another solution $%
n^{2}=n^{2}\left( \mathbf{\alpha }\right) $ must then be obtained from the
backward difference variation as $\left( \Delta ^{b}\right) \Psi \left\{
n^{2}\right\} +\mathbf{\alpha \cdot }\Delta ^{b}\mathbf{\psi }\left\{
n^{2}\right\} =0$. To discriminate the difference between $n^{1}\left( 
\mathbf{\alpha }\right) \ $or $n^{2}\left( \mathbf{\alpha }\right) $, we
examine $\left( \Delta ^{f}\right) ^{2}\Psi \left\{ n^{1}\right\} $ and $%
\left( \Delta ^{b}\right) ^{2}\Psi \left\{ n^{2}\right\} \left( \neq \left(
\Delta ^{f}\right) ^{2}\Psi \left\{ n^{1}\right\} \right) $. Thus, it is
reasonable to conjecture that if $\left\vert \left( \Delta ^{f}\right)
^{2}\Psi \left\{ n^{1}\right\} \right\vert >\left\vert \left( \Delta
^{b}\right) ^{2}\Psi \left\{ n^{2}\right\} \right\vert $, the solution $%
n^{1}=n^{1}\left( \mathbf{\alpha }\right) $ is true, and \textit{vice versa}%
. In other words, the second order difference of the true solution take
largest value as $\max \left\{ \left\vert \Delta ^{2}\Psi \left\{
n^{1}\right\} \right\vert ,\left\vert \Delta ^{2}\Psi \left\{ n^{2}\right\}
\right\vert \right\} $, where $\Delta ^{2}$ must be used in the sense of the 
\emph{unidirectional differences}. This is another point of the new
procedure greatly different from the conventional one that accepts all
solutions once they are stationary.

If the \emph{discrete function} $\Psi \left( n\right) $ is the sum of two IFs%
\emph{\ }$\Psi _{1}\left( n\right) $ and $\Psi _{2}\left( n\right) $, $\Psi
\left( n\right) =\Psi _{1}\left( n\right) +\Psi _{2}\left( n\right) $, there
are four relations $n^{\xi }=n^{\xi }\left( \mathbf{\alpha }\right) $ ($\xi
=1,2,3,4$) from (\ref{variation}) in the solution pool, satisfying,
respectively, 
\begin{subequations}
\begin{eqnarray}
\left( \Delta ^{f}\right) \Psi _{1}\left( n^{1}\right) +\left( \Delta
^{f}\right) \Psi _{2}\left( n^{1}\right) +\mathbf{\alpha \cdot }\Delta 
\mathbf{\psi }\left\{ n^{1}\right\} &=&0,\text{ for 1f2f,}  \label{T21} \\
\left( \Delta ^{f}\right) \Psi _{1}\left( n^{2}\right) +\left( \Delta
^{b}\right) \Psi _{2}\left( n^{2}\right) +\mathbf{\alpha \cdot }\Delta 
\mathbf{\psi }\left\{ n^{2}\right\} &=&0,\text{ for 1f2b,}  \label{T22} \\
\left( \Delta ^{b}\right) \Psi _{1}\left( n^{3}\right) +\left( \Delta
^{f}\right) \Psi _{2}\left( n^{3}\right) +\mathbf{\alpha \cdot }\Delta 
\mathbf{\psi }\left\{ n^{3}\right\} &=&0,\text{ for 1b2f,}  \label{T23} \\
\left( \Delta ^{b}\right) \Psi _{1}\left( n^{4}\right) +\left( \Delta
^{b}\right) \Psi _{2}\left( n^{4}\right) +\mathbf{\alpha \cdot }\Delta 
\mathbf{\psi }\left\{ n^{4}\right\} &=&0,\text{ for 1b2b,}  \label{T24}
\end{eqnarray}%
where the simple form of the constraint functions $\mathbf{\psi }$ all
proportional to $n$ is assumed, for simplicity, so that $\Delta \mathbf{\psi 
}\left\{ n^{\mu }\right\} =\left( \Delta ^{f}\right) \mathbf{\psi }\left\{
n\right\} =\left( \Delta ^{b}\right) \mathbf{\psi }\left\{ n\right\} $. The
mixing form of the solution 1f2b (\ref{T22}) and 1b2f (\ref{T23}) can not be
excluded but has not been noted before. It is understandable, for a
continuous function $f\left( x\right) =f_{1}\left( x\right) +f_{2}\left(
x\right) $, no difference is possible in differentials ($df=df_{1}+$ $df_{2}$%
) or derivatives ($df/dx=df_{1}/dx+df_{2}/dx$), when these operations acting
on two functions $f_{1}\left( x\right) $ and $f_{2}\left( x\right) $; and it
is then reasonably infer that when $f\left( x\right) =f_{1}\left( x\right)
+f_{2}\left( x\right) $, we have still $\Delta f\left( x\right) =\Delta
f_{1}\left( x\right) +\Delta f_{2}\left( x\right) $ without mixing of $%
\Delta ^{f}$ and $\Delta ^{b}$. The new procedure conjectures that the true
solution $n^{\xi }=n^{\xi }\left( \mathbf{\alpha }\right) $ is that its
second order difference $\Delta ^{2}\Psi \left( n^{\xi }\right) $ takes the
largest value in magnitude among all solutions in the solution pool, 
\end{subequations}
\begin{equation}
\left\vert \Delta ^{2}\Psi \left( n^{\xi }\right) \right\vert =\max \left\{
\left\vert \Delta ^{2}\Psi \left\{ n^{1}\right\} \right\vert \text{,}%
\left\vert \Delta ^{2}\Psi \left\{ n^{2}\right\} \right\vert \text{,}%
\left\vert \Delta ^{2}\Psi \left\{ n^{3}\right\} \right\vert \text{,}%
\left\vert \Delta ^{2}\Psi \left\{ n^{4}\right\} \right\vert \right\} .
\label{Sol}
\end{equation}

Five immediate comments follow. 1. The new procedure is essentially a method
of MPD, and importantly during all derivation steps, no one requires that
the variable $n$ be very large. 2. In the new procedure, the correct
solution and the true solution differ; and the former refers to its solving
the first order equation (\ref{variation}) and the latter refers to its
maximizing the second order variations. 3. Once there are degenerate
solutions which have no difference up to second differences, higher order
differences must be invoked. 4. The new procedure can be easily generalized
for the \emph{discrete function} $\Psi \left( n\right) $ that is the sum of
more IFs. 5. If the constraint function $\mathbf{\psi }$ is nonlinear in $n$%
, which is beyond the scope of current studies, the problem must be treated
on the case-by-case base.

We will simply call the new procedure of the discrete calculus of variations
the \emph{IF method}. In the following, I\ will illustrate the IF method
with a detailed derivation of the exact form of the Bose distribution, and
slightly discuss the Boltzmann and Fermi distribution.

\emph{IF method for the Bose system }Considering a system of $N$
noninteracting, indistinguishable particles confined to a space of volume $V$
and sharing a given energy $E$. Let $\varepsilon _{i}$ denote the energy of $%
i$-th level and $\varepsilon _{1}\prec \varepsilon _{2}\prec \varepsilon
_{3}\prec ...$, and $g_{i}$ denote the degeneracy of the level. In a
particular situation, we may have $n_{1}$ particles in the first level $%
\varepsilon _{1}$, $n_{2}$ particles in the second level $\varepsilon _{2}$,
and so on, defining a distribution set $\left\{ n_{i}\right\} $ \cite{phys}.
The number of the distinct microstates in set $\left\{ n_{i}\right\} $ is
then given by,%
\begin{equation}
\Omega \left\{ n_{i}\right\} =\prod_{i}\frac{(n_{i}+g_{i}-1)!}{%
n_{i}!(g_{i}-1)!}.
\end{equation}%
The Bose functional $f$ is,%
\begin{equation}
f=\sum_{i}\left( \ln (n_{i}+g_{i}-1)!-\ln n_{i}!-\ln (g_{i}-1)!\right)
-\alpha \left( \sum_{i}n_{i}-N\right) -\beta \left( \sum_{i}n_{i}\varepsilon
_{i}-E\right) .
\end{equation}%
The variational $\delta f$ is, 
\begin{eqnarray}
\delta f &=&\sum_{i}\left\{ \delta \ln (n_{i}+g_{i}-1)!-\delta \ln
n_{i}!\right\} -\delta n_{i}\left( \alpha +\beta \varepsilon _{i}\right) 
\notag \\
&=&\sum_{i}\delta n_{i}\left\{ \frac{\delta \ln (n_{i}+g_{i}-1)!}{\delta
n_{i}}-\frac{\delta \ln n_{i}!}{\delta n_{i}}-\left( \alpha +\beta
\varepsilon _{i}\right) \right\} .
\end{eqnarray}%
Since the independence of variables $n_{i}$, $\delta f=0$ leads to,%
\begin{equation}
\frac{\delta \ln (n_{i}+g_{i}-1)!}{\delta n_{i}}-\frac{\delta \ln n_{i}!}{%
\delta n_{i}}-\left( \alpha +\beta \varepsilon _{i}\right) =0.
\end{equation}%
Thus, there are essentially \emph{two IFs }$\Psi _{1}=\ln (n_{i}+g_{i}-1)!$
and $\Psi _{2}=\ln n_{i}!$\emph{. }For $\ln (n_{i}+g_{i}-1)!$ and $\ln
n_{i}! $, the smallest forward and backward differences are given in Table (%
\ref{T1}). Four combinations constructing $\delta \ln
(n_{i}+g_{i}-1)!/\delta n_{i}-\delta \ln n_{i}^{\ast }!/\delta n_{i}$ are
summarized in Table\ (\ref{T2}). Accordingly, we have four solutions
presented in Table (\ref{T3}), and all satisfy $\delta f=0$. These
distributions are mutually different when $n_{i}\sim 1$, though all of them
converge to the same one when $n_{i}\gg 1$, 
\begin{equation}
n_{i}\approx \frac{g_{i}}{e^{\alpha +\beta \varepsilon _{i}}-1}.
\end{equation}%
\begin{table}[h]
\caption{The smallest forward/backward difference of $\Psi _{1}=\ln
(n_{i}+g_{i}-1)!$ and $\Psi _{2}=\ln n_{i}!$.}
\label{T1}\centering
\par
\begin{tabular}{|l|l|l|}
\hline
& $\Psi _{1}=\ln (n_{i}+g_{i}-1)!$ & $\Psi _{2}=\ln n_{i}!$ \\ \hline
forward & $\ln (n_{i}+g_{i})$ & $\ln (n_{i}+1)$ \\ \hline
backward & $\ln (n_{i}+g_{i}-1)$ & $\ln (n_{i})$ \\ \hline
\end{tabular}%
\end{table}
\begin{table}[h]
\caption{The column and row give the forward/backward differences for $\Psi
_{1}=\ln (n_{i}+g_{i}-1)!$ and $\Psi _{2}=\ln n_{i}!$, respectively, and we
list the results of four combinations $\Delta \ln (n_{i}+g_{i}-1)!-\Delta
\ln n_{i}!$.}
\label{T2}\centering
\par
\begin{tabular}{|l|l|l|}
\hline
& $\Psi _{2}$ forward & $\Psi _{2}$ backward \\ \hline
$\Psi _{1}$ forward & $\ln (n_{i}+g_{i})-\ln (n_{i}+1)$ for 1f2f & $\ln
(n_{i}+g_{i})-\ln n_{i}$ for 1f2b \\ \hline
$\Psi _{1}$ backward & $\ln (n_{i}+g_{i}-1)-\ln (n_{i}+1)$ for 1b2f & $\ln
(n_{i}+g_{i}-1)-\ln n_{i}$ for 1b2b \\ \hline
\end{tabular}%
\end{table}
\begin{table}[h]
\caption{Four solutions}
\label{T3}\centering
\par
\begin{tabular}{|l|l|l|}
\hline
& $\Psi _{2}$ forward & $\Psi _{2}$ backward \\ \hline
$\Psi _{1}$ forward & $\frac{n_{i}+g_{i}}{n_{i}+1}=\exp \left( \alpha +\beta
\varepsilon _{i}\right) $ for 1f2f & $\frac{n_{i}+g_{i}}{n_{i}}=$ $\exp
\left( \alpha +\beta \varepsilon _{i}\right) $ for 1f2b \\ \hline
$\Psi _{1}$ backward & $\frac{n_{i}+g_{i}-1}{n_{i}+1}=$ $\exp \left( \alpha
+\beta \varepsilon _{i}\right) $ for 1b2f & $\frac{n_{i}+g_{i}-1}{n_{i}}%
=\exp \left( \alpha +\beta \varepsilon _{i}\right) $ for 1b2b \\ \hline
\end{tabular}%
\end{table}
The second order variations of $\delta ^{2}\ln \Omega \left\{
n_{i}{}\right\} $ for the four solutions are explicitly shown in the Table (%
\ref{T4}). 
\begin{table}[h]
\caption{The column and row give the forward/backward finite differences for 
$\ln (n_{i}+g_{i}-1)!$ and $\ln n_{i}!$, respectively, and then forming $%
\Delta ^{2}\ln (n_{i}+g_{i}-1)!-\Delta ^{2}\ln n_{i}!$ accordingly.}
\label{T4}\centering
\par
\begin{tabular}{|l|l|l|}
\hline
& $\Psi _{2}$ forward & $\Psi _{2}$ backward \\ \hline
$\Psi _{1}$ forward & $-\ln (\frac{n_{i}+2}{n_{i}+1}\frac{n_{i}+g_{i}}{%
n_{i}+g_{i}+1})$\ for 1f2f & $-\ln (\frac{n_{i}}{n_{i}-1}\frac{n_{i}+g_{i}}{%
n_{i}+g_{i}+1})$\ for 1f2b \\ \hline
$\Psi _{1}$ backward & $-\ln (\frac{n_{i}+2}{n_{i}+1}\frac{n_{i}+g_{i}-2}{%
n_{i}+g_{i}-1})$ for 1b2f & $-\ln (\frac{n_{i}}{n_{i}-1}\frac{n_{i}+g_{i}-2}{%
n_{i}+g_{i}-1})$\ for 1b2b \\ \hline
\end{tabular}%
\end{table}
In order to eliminate the spurious solutions, we examine which one is the
largest in magnitude. It is easily to verify that one combination 1f2b is
the only right one for we have, 
\begin{subequations}
\begin{eqnarray}
\frac{n_{i}}{n_{i}-1}\frac{n_{i}+g_{i}}{n_{i}+g_{i}+1}-\frac{n_{i}}{n_{i}-1}%
\frac{n_{i}+g_{i}-2}{n_{i}+g_{i}-1} &=&\frac{n_{i}}{n_{i}-1}\left( \frac{2}{%
\left( n_{i}+g_{i}\right) ^{2}-1}\right) >0,\text{ }\left( n_{i}>1\right)
\label{Bose1} \\
\frac{n_{i}}{n_{i}-1}\frac{n_{i}+g_{i}}{n_{i}+g_{i}+1}-\frac{n_{i}+2}{n_{i}+1%
}\frac{n_{i}+g_{i}}{n_{i}+g_{i}+1} &=&\frac{n_{i}+g_{i}}{n_{i}+g_{i}+1}%
\left( \frac{2}{n_{i}{}^{2}-1}\right) >0,\text{ }\left( n_{i}>1\right)
\label{Bose2} \\
\frac{n_{i}}{n_{i}-1}\frac{n_{i}+g_{i}}{n_{i}+g_{i}+1}-\frac{n_{i}+2}{n_{i}+1%
}\frac{n_{i}+g_{i}-2}{n_{i}+g_{i}-1} &=&2\frac{%
(2n_{i}{}^{2}+2g_{i}n_{i}-2-g_{i}+g_{i}^{2})}{\left( \left(
n_{i}+g_{i}\right) ^{2}-1\right) \left( n_{i}{}^{2}-1\right) }>0,\text{ }%
\left( n_{i}>1\right) .  \label{Bose3}
\end{eqnarray}%
To note that once $n_{i}=0$, and $n_{i}=1$, we need to directly invoke the
expression for 1f2b in Table (\ref{T4}) and the $\left\vert \ln
(n_{i}/(n_{i}-1)\right\vert \rightarrow \infty $, and in final we obtain the
Bose distribution, 
\end{subequations}
\begin{equation}
n_{i}=\frac{g_{i}}{e^{\alpha +\beta \varepsilon _{i}}-1}.  \label{BD}
\end{equation}

Two remarks follow. 1. We can follow the similar manner to present detailed
derivations for Boltzmann and Fermi distribution. For the Boltzmann
distribution, there is essentially one IF and the true solution comes from
backward difference; ; and for the Fermi distribution there are essentially
two IFs and the true solution comes from the combination 1b2b. 2, Our
procedure puts no requirement on the total number of particle $N$ that be
very large. Within the grand ensemble theory, we can easily show the Bose
distribution to be given by with $g_{i}=1$, 
\begin{equation}
n_{i}=\frac{1}{e^{\alpha +\beta \varepsilon _{i}}-1}+\frac{1+N}{e^{\left(
\alpha +\beta \varepsilon _{i}\right) \left( 1+N\right) }-1}.
\end{equation}%
It differs from ours (\ref{BD}) by an additional term depending on $N$.

\emph{Conclusions and discussions }In contrast to the continuous calculus of
variations, the discrete one possesses some peculiarities. The function in
it can not be treated in the uniformly forward or backward difference, but
must exhaust all mathematical possibilities. All possible solutions
determined by the first order variation equation include both true and false
one. Each of the solutions has its own second order variation and the true
solution takes the largest value in magnitude. The application of the
procedure to the statistical distributions is successful, giving results
identical to those obtained from the grand ensemble theory. However, the
ensemble theory is valid only when the total number of particle in the
system is large. Our approach does not suffer from such a limitation.

\begin{acknowledgments}
This work is financially supported by National Natural Science Foundation of
China under Grant No. 11675051.
\end{acknowledgments}

\end{document}